\documentclass[conference]{IEEEtran}
\IEEEoverridecommandlockouts

\usepackage{cite}
\usepackage{amsmath,amssymb,amsfonts}
\usepackage{algorithmic}
\usepackage{graphicx}
\usepackage{textcomp}
\usepackage{xcolor}
\usepackage{braket}
\usepackage{tikz}
\usepackage{qcircuit}
\usepackage{lmodern}     
\usepackage[T1]{fontenc}  
\usepackage{atbegshi}
\usepackage{eso-pic}
\AtBeginShipout{%
  \ifnum\value{page}=4
\global\setbox\AtBeginShipoutBox=\vbox{\vspace{2pt}\unvbox\AtBeginShipoutBox}%
  \fi
}
\newcommand{\RY}[1]{\operatorname{RY}\left(#1\right)}
\newcommand{\RZ}[1]{\operatorname{RZ}\left(#1\right)}
\newcommand{\CNOT}{\textsc{CNOT}}
\newcommand{\ketzero}{\ket{0}^{\otimes n}}

\usetikzlibrary{positioning, calc, shapes.geometric, arrows.meta}

\def\BibTeX{{\rm B\kern-.05em{\sc i\kern-.025em b}\kern-.08em
    T\kern-.1667em\lower.7ex\hbox{E}\kern-.125emX}}
\begin{document}

\title{Q-DPTS: Quantum Differentially Private Time Series Forecasting via Variational Quantum Circuits\thanks{The views expressed in this article are those of the authors and do not represent the views of Wells Fargo. This article is for informational purposes only. Nothing contained in this article should be construed as investment advice. Wells Fargo makes no express or implied warranties and expressly disclaims all legal, tax, and accounting implications related to this article. }\\

}

\author{
\IEEEauthorblockN{Chi-Sheng Chen}
\IEEEauthorblockA{\textit{Independent Researcher} \\
Cambridge, USA \\
m50816m50816@gmail.com}
\and
\IEEEauthorblockN{Samuel Yen-Chi Chen}
\IEEEauthorblockA{\textit{Wells Fargo} \\
New York, USA \\
yen-chi.chen@wellsfargo.com}
}

\maketitle

\begin{abstract}
Time series forecasting is vital in domains where data sensitivity is paramount, such as finance and energy systems. While Differential Privacy (DP) provides theoretical guarantees to protect individual data contributions, its integration especially via DP-SGD often impairs model performance due to injected noise. In this paper, we propose Q-DPTS, a hybrid quantum-classical framework for Quantum Differentially Private Time Series Forecasting. Q-DPTS combines Variational Quantum Circuits (VQCs) with per-sample gradient clipping and Gaussian noise injection, ensuring rigorous $(\epsilon, \delta)$-differential privacy. The expressiveness of quantum models enables improved robustness against the utility loss induced by DP mechanisms. We evaluate Q-DPTS on the ETT (Electricity Transformer Temperature) dataset, a standard benchmark for long-term time series forecasting. Our approach is compared against both classical and quantum baselines, including LSTM, QASA, QRWKV, and QLSTM. Results demonstrate that Q-DPTS consistently achieves lower prediction error under the same privacy budget, indicating a favorable privacy-utility trade-off. This work presents one of the first explorations into quantum-enhanced differentially private forecasting, offering promising directions for secure and accurate time series modeling in privacy-critical scenarios.
\end{abstract}


\section{Introduction}
The rapid growth of data-driven applications in healthcare, finance, and smart infrastructure has highlighted the dual need for powerful predictive models and rigorous privacy guarantees. Time series data, in particular, often contain sensitive personal or organizational information, making the integration of privacy-preserving mechanisms into forecasting algorithms a pressing concern.

Differential Privacy (DP) has emerged as a principled approach to ensure individual-level data protection by injecting controlled noise into the learning process \cite{dwork2006differential}. In parallel, Quantum Machine Learning (QML) has garnered attention for its potential to outperform classical models in specific learning regimes due to richer representational capacity and quantum-enhanced optimization \cite{mitarai2018quantum}. However, despite their respective advances, the intersection of time-series data on quantum models and differential privacy remains largely unexplored, particularly in the domain of temporal data modeling.

\begin{figure}
    \centering
    \includegraphics[width=1\linewidth]{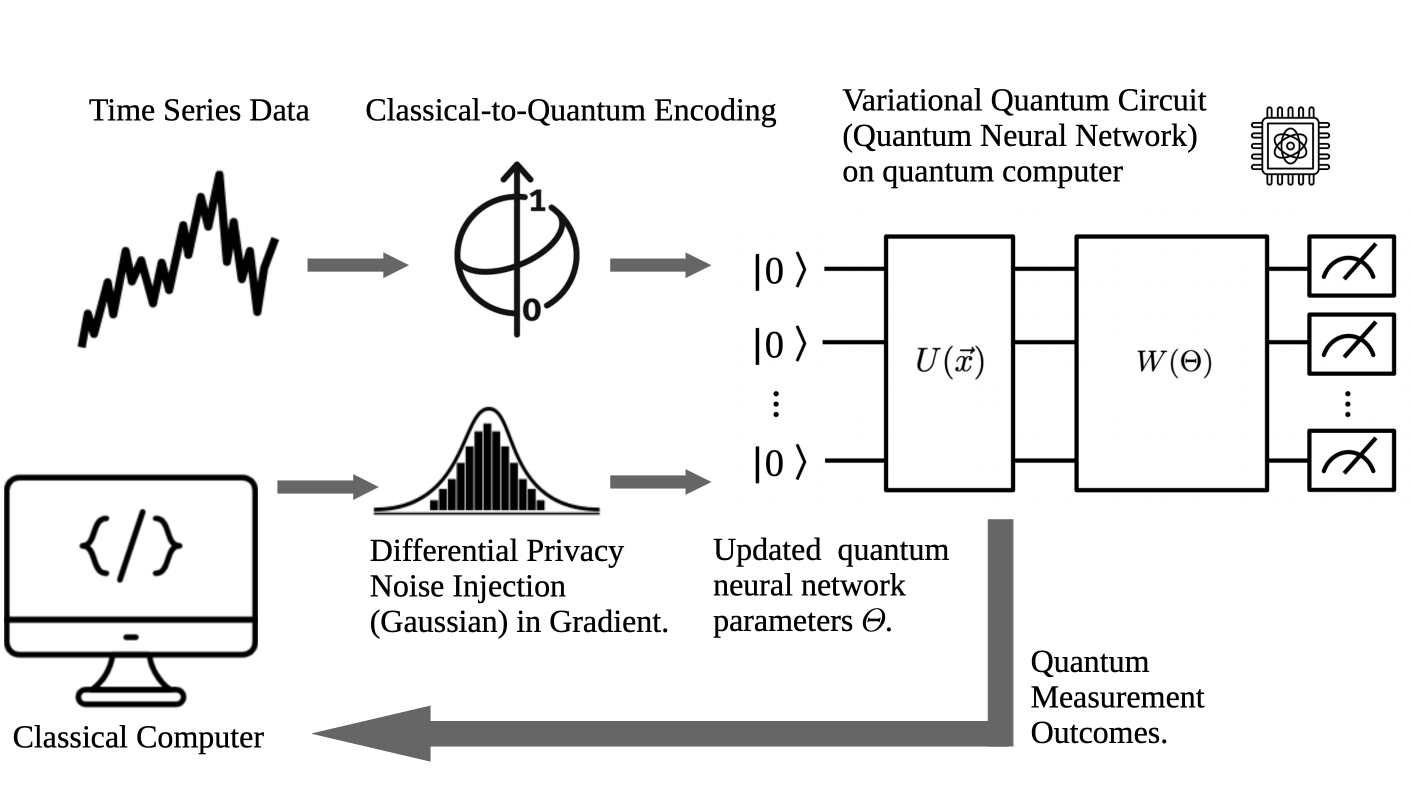}
    \caption{The overview of this work.}
    \label{fig:overview}
\end{figure}

In this work, we present \textbf{Q-DPTS}, as shown in Fig.~\ref{fig:overview}, the first quantum forecasting framework that satisfies differential privacy. Q-DPTS leverages variational quantum circuits (VQCs) as core computational modules for time series modeling, while applying Rényi Differential Privacy (RDP) accounting \cite{mironov2017renyi} and Gaussian mechanism-based gradient perturbation during training. This hybrid approach combines the generalization capabilities of quantum models with provable privacy guarantees. But why DP + Quantum? DP-SGD often hurts utility due to gradient noise. Hybrid quantum models offer strong function classes via entanglement and measurement aggregation, which can compensate for DP-induced noise especially in short-horizon forecasting. Moreover, shot-based estimation naturally regularizes updates, and per-sample clipping aligns well with the bounded-observable structure of VQCs. Our hypothesis is that quantum expressivity plus DP smoothing yields competitive privacy-utility trade-offs in temporal data, which we test in this paper.

We implement and evaluate four quantum architectures: Quantum Long Short-Term Memory (QLSTM) \cite{chen2022qlstm}, Quantum Recurrent Weighted Key-Value (QRWKV) \cite{chen2025qrwkv}, and Quantum Adaptive Self-Attention (QASA) \cite{chen2025qasa}. Each model is trained under varying noise multipliers corresponding to different privacy budgets ($\sigma = 0, 0.5, 1.0, 2.0$), and is assessed using mean absolute error (MAE), mean squared error (MSE), and root mean squared error (RMSE) on the ETT dataset \cite{zhou2021informer}, a benchmark for multivariate time series forecasting.

To the best of our knowledge, this is the first work to explore differentially private training of variational quantum models for time series forecasting. Our contributions are threefold:
\begin{itemize}
    \item We propose Q-DPTS, the first framework to integrate differential privacy into quantum forecasting models.
    \item We design and implement a suite of hybrid quantum-classical architectures compatible with DP-SGD training using RDP-based noise accounting.
    \item We benchmark model performance under various privacy levels, revealing that certain quantum models (notably QASA and QLSTM) demonstrate robustness to noise and maintain competitive forecasting accuracy.
\end{itemize}

Our findings suggest that differentially private quantum forecasting is not only feasible but also effective, offering a promising direction for privacy-preserving modeling in quantum-enhanced environments.

\section{Related Works}

\subsection{Quantum Time Series Forecasting}
Quantum machine learning (QML) \cite{biamonte2017quantum} has recently been applied to time series modeling, aiming to leverage the high expressivity and entanglement properties of parameterized quantum circuits (PQCs). Hybrid quantum-classical models such as Quantum Long Short-Term Memory (QLSTM) and variational quantum circuits (VQCs) have demonstrated the ability to capture complex temporal dependencies \cite{khan2024quantum}. Other architectures, including quantum-enhanced attention mechanisms and Fourier-based encoders, have also been proposed for forecasting and anomaly detection in sequential datasets \cite{liu2018quantum}. However, existing quantum models are typically evaluated under noise-free hardware simulations (i.e., no gate/channel noise in the quantum backend), assuming full access to raw training data, which limits their real-world applicability in privacy-sensitive domains.

\subsection{Differential Privacy in Time Series Forecasting}
Differential privacy (DP) has been widely adopted as a formal mechanism to protect individual data points in machine learning workflows. In time series forecasting, DP-SGD \cite{du2021dynamic} and Rényi Differential Privacy (RDP) accounting have been successfully integrated into deep learning models to limit the privacy leakage of sensitive sequences. Recent efforts have explored differentially private recurrent networks \cite{mcmahan2017learning}, federated time series prediction \cite{yang2023differentially}, and privacy-preserving sensor modeling. However, these techniques have only been applied to classical architectures, and their extension to quantum models has not yet been realized.

\subsection{Quantum Differential Privacy}
The intersection of quantum computing and differential privacy remains largely theoretical. Early work in quantum cryptography has explored the definition of quantum differential privacy (QDP) in the context of quantum data and queries \cite{zhou2017differential}. More recent studies have proposed privacy guarantees for quantum algorithms via measurement noise or randomized response techniques \cite{watkins2023quantum}. Nonetheless, no existing research has implemented DP-SGD within variational quantum learning pipelines for practical machine learning tasks such as forecasting. Our work fills this gap by being the first to empirically evaluate the effect of DP-induced noise on quantum forecasting models and to benchmark multiple quantum architectures under different privacy budgets.

\section{Methodology}

We present four quantum time series forecasting models built on hybrid quantum-classical architectures. Each model encodes a fixed-length historical window of $T=16$ time steps of multivariate data $\mathbf{X} = [\mathbf{x}_{t-T+1}, \dots, \mathbf{x}_t] \in \mathbb{R}^{T \times d}$ into a quantum representation, processes it with variational quantum circuits (VQCs), and predicts the value $\hat{y}_{t+1}$ of the next time step. This setting follows a 16-step input $\rightarrow$ 1-step output formulation, distinct from sequence-to-sequence models.

All models support differentially private training using DP-SGD. At each training step, gradients are clipped to a fixed $\ell_2$ norm bound $C$, and Gaussian noise $\mathcal{N}(0, \sigma^2 C^2)$ is added before averaging and updating the parameters. Privacy guarantees are tracked using Rényi Differential Privacy (RDP) accounting, where the total privacy loss $\epsilon$ is computed as a function of noise multiplier $\sigma$, number of steps $T$, sampling rate $\beta$, and target $\delta$:

\begin{equation}
\epsilon(\sigma, T, \beta, \delta) \leq \min_{\alpha > 1} \left( \frac{1}{\alpha - 1} \log \left( \frac{1}{\delta} \right) + \frac{\alpha \beta^2 T}{2 \sigma^2} \right).
\end{equation}

\subsection{Why DP-QML Can Be Robust (Theory Sketch)}

Consider one training step of DP-SGD with per-sample clipping at norm $C$ and Gaussian noise 
$\mathcal{N}(0,\sigma^2 C^2 I)$. 
For a variational quantum circuit (VQC) with measurement observable 
$O = \sum_i w_i Z_i$ and depth-$L$ circuit $U(\theta)$, 
the prediction is
\begin{equation}
    f_\theta(x) = \langle 0 | U^\dagger(\theta) O U(\theta) | 0 \rangle .
\end{equation}
This mapping is $L$-Lipschitz in $\theta$ under bounded operator norms. 
The parameter-shift rule provides an unbiased per-sample gradient estimator. 
With clipping, the sensitivity is bounded:
\begin{equation}
    \| \nabla_\theta \ell(f_\theta(x), y) \|_2 \le C .
\end{equation}
The DP noise then adds a zero-mean perturbation with variance $\sigma^2 C^2$, 
which is equivalent to smoothing the loss landscape. 

In hybrid models (QASA, QRWKV, QLSTM), quantum measurements average over qubits or timesteps, 
further reducing gradient variance. 
Combining: (i) Lipschitz outputs of shallow VQCs, (ii) clipping-bounded sensitivity, and 
(iii) Gaussian smoothing implied by the Rényi DP bound in Eq.~(1), 
we obtain a smaller excess-risk term of order $\mathcal{O}(\sigma^2 C^2)$ while preserving expressivity. 
This provides a mechanism-level explanation for the empirical robustness observed in Table~\ref{tab:results} (Sec.~V).

\paragraph{Remark.}
Under standard smoothness assumptions and a stepsize schedule with 
$\sum_t \eta_t = \infty$ and $\sum_t \eta_t^2 < \infty$, 
DP-SGD with parameter-shift remains convergent to stationary points in expectation. 
The DP noise increases gradient variance but does not bias the estimator.

We describe each quantum model below, standardizing their structure into: (1) classical-to-quantum encoding, (2) VQC processing, and (3) measurement and decoding.

\begin{figure}
    \centering
    \includegraphics[width=1\linewidth]{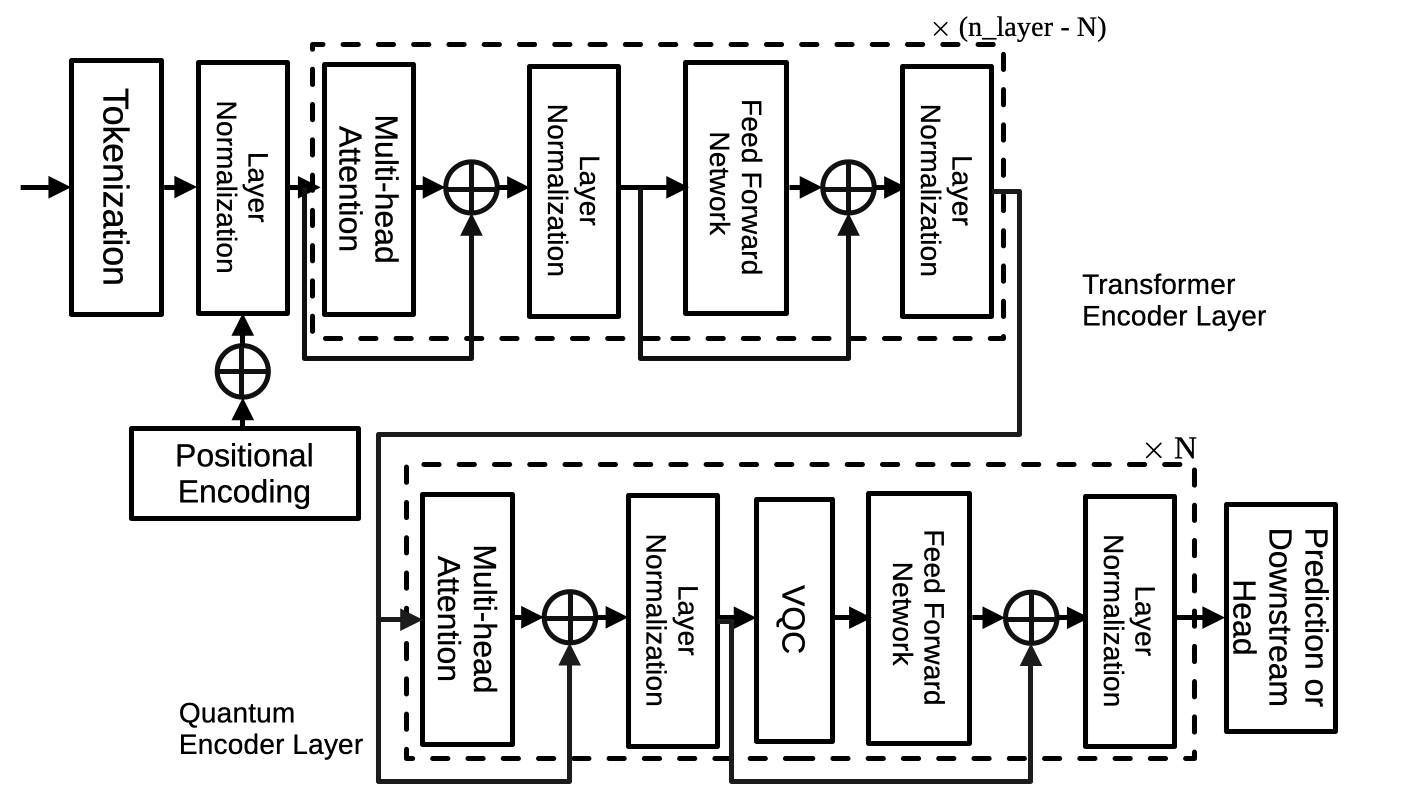}
    \caption{The model architecture of QASA.}
    \label{fig:qasa}
\end{figure}

\begin{figure}
    \centering
    \includegraphics[width=1\linewidth]{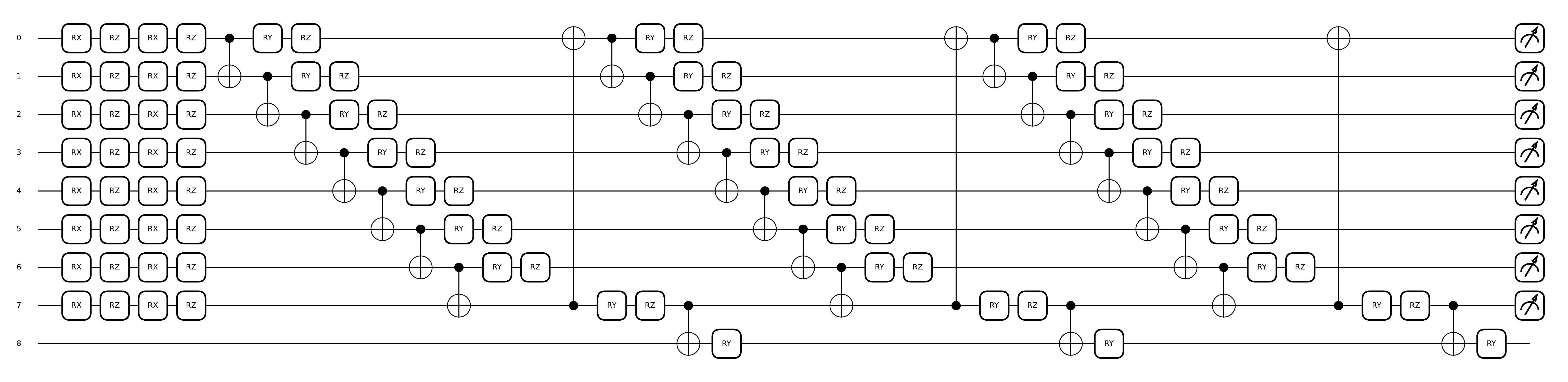}
    \caption{The VQC used in QASA.}
    \label{fig:qasa_vqc}
\end{figure}

\subsection{Quantum Attention Self-Attention (QASA)}

QASA \cite{chen2025qasa} implements self-attention via VQCs, the model as shown in Fig.~\ref{fig:qasa}. Given input sequence $\mathbf{X} \in \mathbb{R}^{T \times d}$, we compute quantum query, key, and value embeddings:

\begin{equation}
\mathbf{Q}_t = \text{VQC}_q(\mathbf{x}_t),\quad
\mathbf{K}_t = \text{VQC}_k(\mathbf{x}_t),\quad
\mathbf{V}_t = \text{VQC}_v(\mathbf{x}_t).
\end{equation}

Each $\text{VQC}(\cdot)$ encodes the vector $\mathbf{x}_t$ into quantum amplitudes, applies a variational quantum circuit composed of $L$ layers of $RY(\theta)$ rotations and entangling CNOT gates, and measures expectation values $\langle Z_i \rangle$ on $n$ qubits like Fig.~\ref{fig:qasa_vqc},:

\begin{equation}
\text{VQC}(\mathbf{x}) = \left( \langle Z_1 \rangle, \langle Z_2 \rangle, \dots, \langle Z_n \rangle \right).
\end{equation}

Each token vector $\mathbf{x}_t \in \mathbb{R}^{d}$ is embedded and encoded as an amplitude-encoded quantum state:

\begin{equation}
    \ket{\psi_t} = \frac{1}{\|\mathbf{x}_t\|} \sum_{i=1}^{d} x_{t,i} \ket{i}.
\end{equation}

Then a variational quantum circuit $U(\boldsymbol{\theta})$ is applied:

\begin{equation}
    U(\boldsymbol{\theta}) = \prod_{\ell=1}^{L} \left[
        \bigotimes_{i=1}^{n} \RY{\theta_{\ell,i}} \cdot \CNOT_{i,i+1}
    \right],
\end{equation}

where $L$ is the number of layers and $n = \lceil \log_2 d \rceil$ qubits are used. The output $\mathbf{z}_t$ is obtained by measuring the expectation values of Pauli-Z operators:

\begin{equation}
    \mathbf{z}_t = \left( \langle Z_1 \rangle, \dots, \langle Z_n \rangle \right).
\end{equation}

Attention~\cite{vaswani2017attention} is computed as:

\begin{equation}
\text{Attention}(\mathbf{Q}, \mathbf{K}, \mathbf{V}) = \text{softmax}\left( \frac{\mathbf{Q} \mathbf{K}^\top}{\sqrt{d}} \right) \mathbf{V}.
\end{equation}

The attention output is decoded through a classical feedforward network to predict $\hat{y}_{t+1}$.

\begin{figure}
    \centering
    \includegraphics[width=1\linewidth]{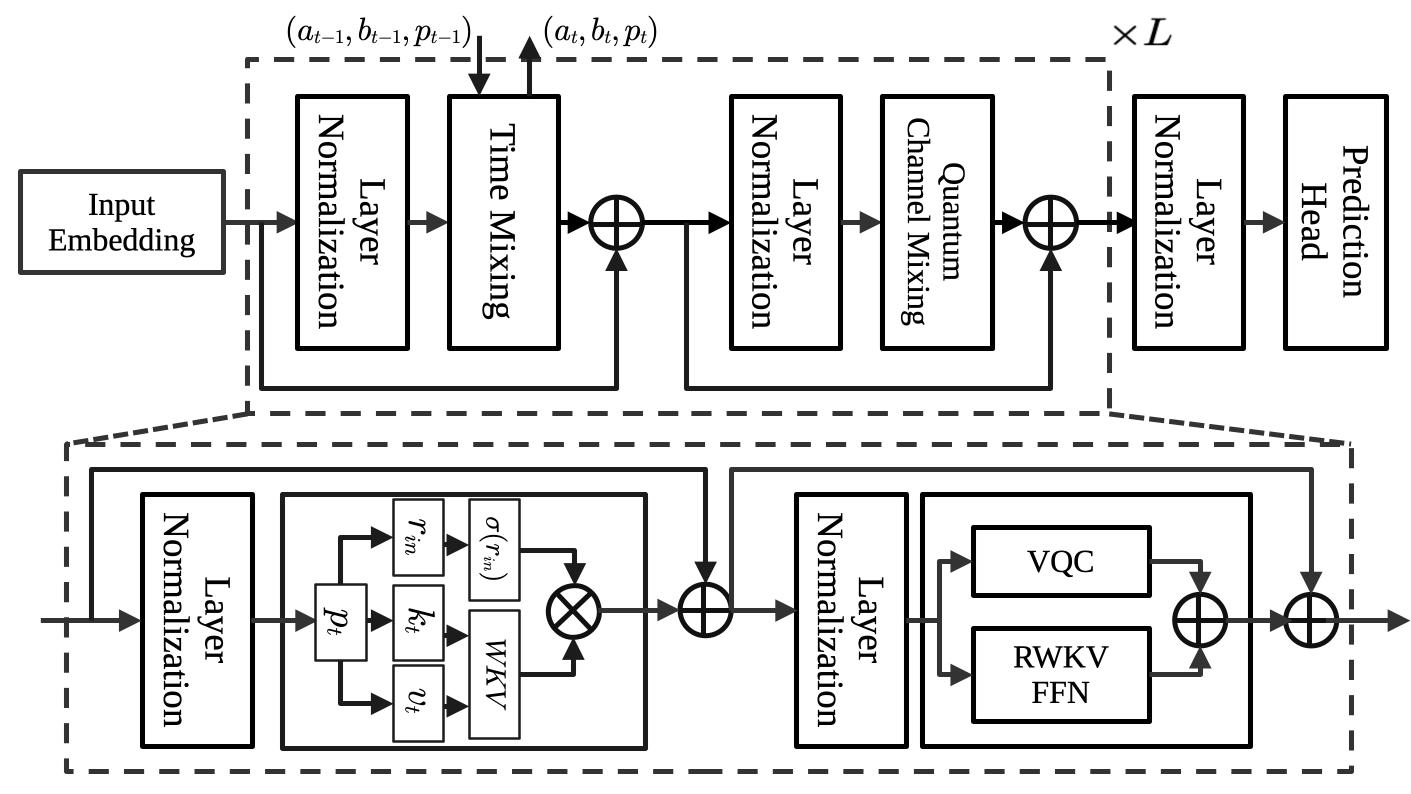}
    \caption{Quantum RWKV architecture.}
    \label{fig:qrwkv}
\end{figure}

\begin{figure}[htbp]
    \centering
    \scalebox{0.8}{ 
    \Qcircuit @C=1em @R=1em {
    & \lstick{\ket{q_0}} & \gate{RX} & \gate{RX} & \ctrl{1} & \qw      & \qw      & \targ     & \gate{RX} & \ctrl{1} & \qw      & \qw      & \targ       & \meter \\
    & \lstick{\ket{q_1}} & \gate{RX} & \gate{RX} & \targ    & \ctrl{1} & \qw      & \qw       & \gate{RX} & \targ    & \ctrl{1} & \qw      & \qw      & \meter \\
    & \lstick{\ket{q_2}} & \gate{RX} & \gate{RX} & \qw      & \targ    & \ctrl{1} & \qw       & \gate{RX} & \qw      & \targ    & \ctrl{1} & \qw      & \meter \\
    & \lstick{\ket{q_3}} & \gate{RX} & \gate{RX} & \qw      & \qw      & \targ    & \ctrl{-3} & \gate{RX} & \qw      & \qw      & \targ    & \ctrl{-3} & \meter
    }
    }
    \caption{The VQC used in Quantum RWKV.}
    \label{fig:qrwkv-vqc}
\end{figure}
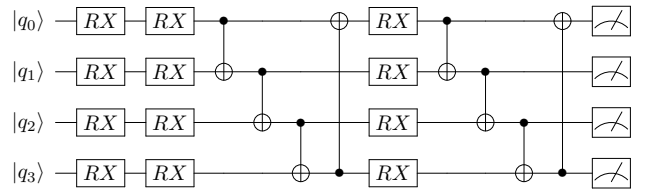

\subsection{Quantum Receptance Weighted Key-Value (QRWKV)}

QRWKV~\cite{chen2025qrwkv} integrates quantum evolution with the receptance attention-free model~\cite{peng2023rwkv}, the model detail is in the Fig.~\ref{fig:qrwkv}. At each time step \(t\), the input \(\mathbf{x}_t\) is first passed through a Variational Quantum Circuit (VQC) to produce a quantum embedding:
\begin{equation}
    \mathbf{h}_t = \mathrm{VQC}_q(\mathbf{x}_t), 
    \quad 
    [\mathbf{q}_t,\;\mathbf{k}_t,\;\mathbf{v}_t] \subseteq \mathbf{h}_t.
\end{equation}
Concretely, we prepare \(\ket{0}^{\otimes n}\) and apply a parameterized circuit \(\;U_\Theta=\prod_{\ell=1}^L U^{(\ell)}\;\) where
\begin{equation}
    U^{(\ell)} = \Bigl(\prod_{i=1}^{n} \RY{\theta^{(\ell)}_{i}} \RZ{\phi^{(\ell)}_{i}}\Bigr)\;\mathrm{EntangleLayer},
\end{equation}
and \(\mathrm{EntangleLayer}\) applies CNOT gates in a chosen pattern. Measurement yields the vector \(\mathbf{h}_t\), from which we split out query \(\mathbf{q}_t\), key \(\mathbf{k}_t\), and value \(\mathbf{v}_t\) sub-vectors.

\paragraph{Time-Mixing and Receptance Gate}  
Classical time-mixing follows the RWKV design. Project the same input \(\mathbf{x}_t\) to key and value signals:
\begin{equation}
\mathbf{u}_t = W^K \mathbf{x}_t,\quad
\mathbf{v}_t = W^V \mathbf{x}_t,
\end{equation}
and accumulate with exponential decay:
\begin{equation}
    \mathbf{m}_t = \lambda\,\mathbf{m}_{t-1} + \mathbf{v}_t,
    \quad 
    \lambda = \exp(-\Delta t / \tau).
\end{equation}
A receptance gate controls the exposed memory:
\begin{equation}
    \mathbf{r}_t = \sigma\bigl(W^R[\mathbf{x}_t;\,\mathbf{m}_{t-1}]+\mathbf{b}^R\bigr),
    \quad W^R\in\mathbb{R}^{d\times2d}.
\end{equation}
The time-mixed output is then
\begin{equation}
    \hat{\mathbf{y}}^{\mathrm{time}}_t = \mathbf{r}_t \odot (\mathbf{u}_t \odot \mathbf{m}_t).
\end{equation}

\paragraph{VQC-Enhanced Channel-Mixing}  
Instead of the classical MLP input, we feed \(\mathbf{x}_t\) into the same VQC to get \(\mathbf{qemb}_t=\mathbf{h}_t\), the VQC detail is in Fig.~\ref{fig:qrwkv-vqc}. The channel-mixing block becomes:
\begin{align}
    \mathbf{z}_t &= W^1\,\mathbf{qemb}_t + W^2\mathrm{MLP}+ \mathbf{b}^1,\\
    \mathbf{h}'_t &= \mathrm{GELU}(\mathbf{z}_t),\\
    \mathbf{c}_t &= W^3\bigl(\mathbf{h}'_t \odot \mathbf{h}'_{t-1}\bigr) + \mathbf{b}^2,
\end{align}
with \(W^1,W^2,W^3\in\mathbb{R}^{d\times d}\) and biases in \(\mathbb{R}^d\). Optionally add residual connections and LayerNorm.

\paragraph{Attention over Quantum Queries and Keys}  
We also compute a measurement-based attention score between quantum-derived queries and keys:
\begin{equation}
    \alpha_{t,\tau}
    = \frac{\exp\bigl\langle \mathbf{q}_t,\mathbf{k}_\tau\bigr\rangle}
           {\sum_{\tau'=1}^t \exp\bigl\langle \mathbf{q}_t,\mathbf{k}_{\tau'}\bigr\rangle},
\end{equation}
and form the attention output
\begin{equation}
    \hat{\mathbf{y}}^{\mathrm{attn}}_{t+1}
    = \sum_{\tau=1}^t \alpha_{t,\tau}\,\mathbf{v}_\tau.
\end{equation}

\paragraph{Full Layer Update}  
Each layer concatenates time-mixing and VQC-enhanced channel-mixing with residuals and normalization:
\begin{equation}
\begin{aligned}
    \mathbf{h}_t &= \mathrm{LayerNorm}\bigl(\mathbf{x}_t + \hat{\mathbf{y}}^{\mathrm{time}}_t\bigr),\\
    \mathbf{y}_t &= \mathrm{LayerNorm}\bigl(\mathbf{h}_t + \mathbf{c}_t + \hat{\mathbf{y}}^{\mathrm{attn}}_t\bigr).
\end{aligned}
\end{equation}


\subsection{Quantum Long Short-Term Memory (QLSTM)}
\begin{figure}[htbp]
\begin{center}
\includegraphics[width=1\columnwidth]{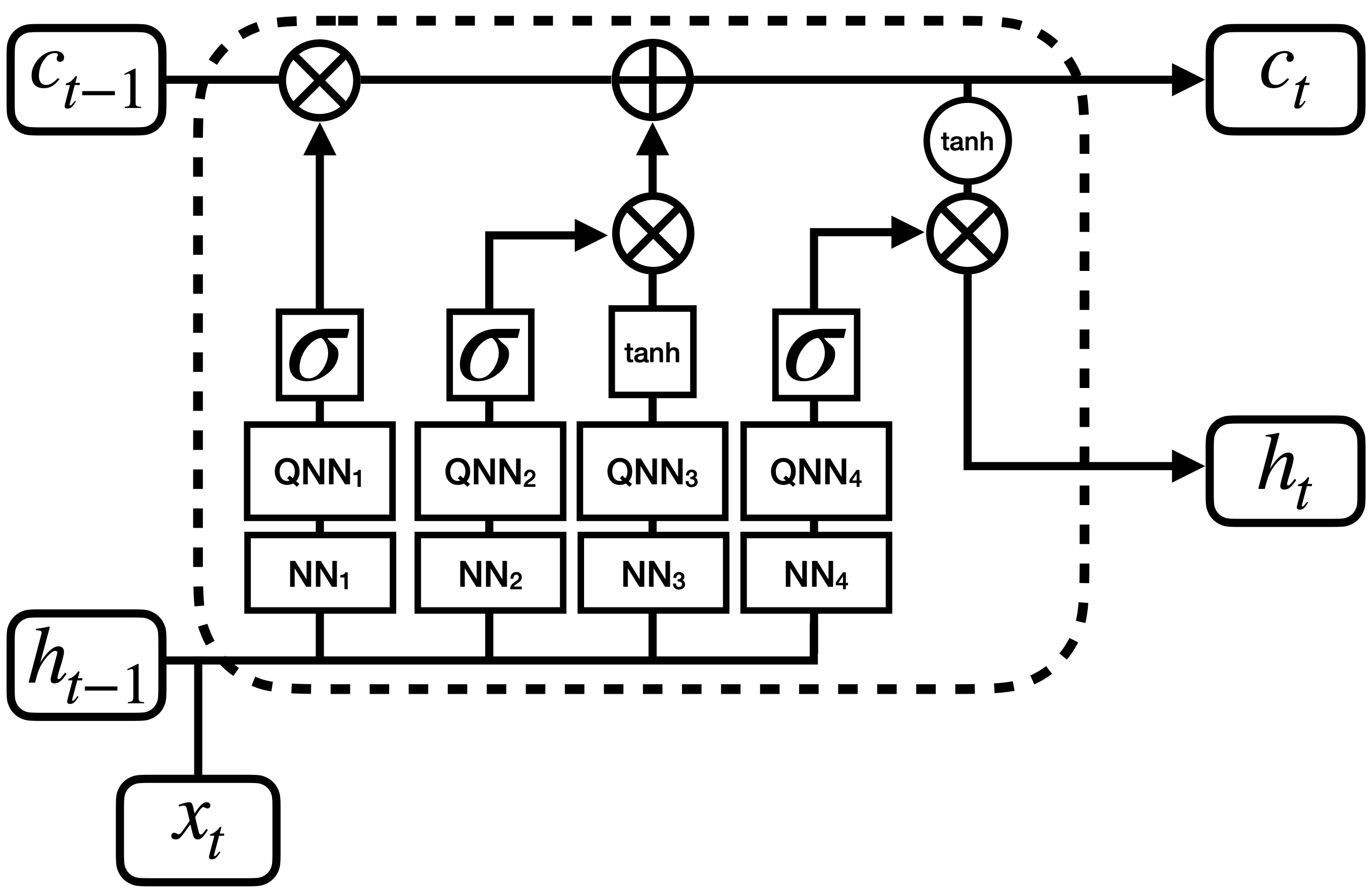}
\caption{Quantum LSTM architecture.}
\label{fig:qlstm}
\end{center}
\end{figure}

QLSTM \cite{chen2022qlstm} adapts classical LSTM \cite{hochreiter1997long} by replacing the classical neural network in each gate with a quantum neural network. A further improved version of QLSTM called linear-enhanced QLSTM \cite{cao2023linear} includes a classical neural network to preprocess the inputs before sending into QNNs. Such modifications have been shown to enhance the learning ability of QLSTM models \cite{cao2023linear}. We adopt this improvement in this benchmark as illustrated in Fig.\ref{fig:qlstm}. Given input $\mathbf{x}_t$ and hidden state $\mathbf{h}_{t-1}$, each gate is computed as:


\begin{align}
\mathbf{f}_t &= \sigma(\text{QNN}_f(\text{NN}_f([\mathbf{x}_t; \mathbf{h}_{t-1}]))), \\
\mathbf{i}_t &= \sigma(\text{QNN}_i(\text{NN}_i([\mathbf{x}_t; \mathbf{h}_{t-1}]))), \\
\mathbf{o}_t &= \sigma(\text{QNN}_o(\text{NN}_o([\mathbf{x}_t; \mathbf{h}_{t-1}]))), \\
\mathbf{g}_t &= \tanh(\text{QNN}_g(\text{NN}_g([\mathbf{x}_t; \mathbf{h}_{t-1}]))).
\end{align}

\begin{figure}[htbp]
\begin{center}
\scalebox{0.7}{
\begin{minipage}{1\columnwidth}
\Qcircuit @C=1em @R=1em {
\lstick{\ket{0}} & \gate{H} & \gate{RY(\arctan(x_1))} & \gate{RZ(\arctan(x_1^2))}       & \ctrl{1}   & \qw       & \gate{RY(\alpha_1)} & \meter \qw \\
\lstick{\ket{0}} & \gate{H} & \gate{RY(\arctan(x_2))} & \gate{RZ(\arctan(x_2^2))}       & \targ      & \ctrl{1}  & \gate{RY(\alpha_2)} & \meter \qw \\
\lstick{\ket{0}} & \gate{H} & \gate{RY(\arctan(x_3))} & \gate{RZ(\arctan(x_3^2))}       & \ctrl{1}   & \targ     & \gate{RY(\alpha_3)} & \meter \qw \\
\lstick{\ket{0}} & \gate{H} & \gate{RY(\arctan(x_4))} & \gate{RZ(\arctan(x_4^2))}       & \targ      & \ctrl{1}  & \gate{RY(\alpha_4)} & \meter \qw \\
\lstick{\ket{0}} & \gate{H} & \gate{RY(\arctan(x_5))} & \gate{RZ(\arctan(x_5^2))}       & \ctrl{1}   & \targ     & \gate{RY(\alpha_5)} & \meter \qw \\
\lstick{\ket{0}} & \gate{H} & \gate{RY(\arctan(x_6))} & \gate{RZ(\arctan(x_6^2))}       & \targ      & \ctrl{1}  & \gate{RY(\alpha_6)} & \meter \qw \\
\lstick{\ket{0}} & \gate{H} & \gate{RY(\arctan(x_7))} & \gate{RZ(\arctan(x_7^2))}       & \ctrl{1}   & \targ     & \gate{RY(\alpha_7)} & \meter \qw \\
\lstick{\ket{0}} & \gate{H} & \gate{RY(\arctan(x_8))} & \gate{RZ(\arctan(x_8^2))}       & \targ      & \ctrl{1}  & \gate{RY(\alpha_8)} & \meter \qw \\
\lstick{\ket{0}} & \gate{H} & \gate{RY(\arctan(x_9))} & \gate{RZ(\arctan(x_9^2))}       & \ctrl{1}   & \targ     & \gate{RY(\alpha_9)} & \meter \qw \\
\lstick{\ket{0}} & \gate{H} & \gate{RY(\arctan(x_{10}))} & \gate{RZ(\arctan(x_{10}^2))} & \targ      & \ctrl{1}  & \gate{RY(\alpha_{10})} & \meter \qw \\
\lstick{\ket{0}} & \gate{H} & \gate{RY(\arctan(x_{11}))} & \gate{RZ(\arctan(x_{11}^2))} & \qw        & \targ     & \gate{RY(\alpha_{11})}& \meter \gategroup{1}{5}{11}{7}{.7em}{--}\qw 
}
\end{minipage}}
\end{center}
\caption{Generic VQC architecture for QLSTM.}
\label{Fig:QLSTM_VQC_CIRCUIT}
\end{figure}
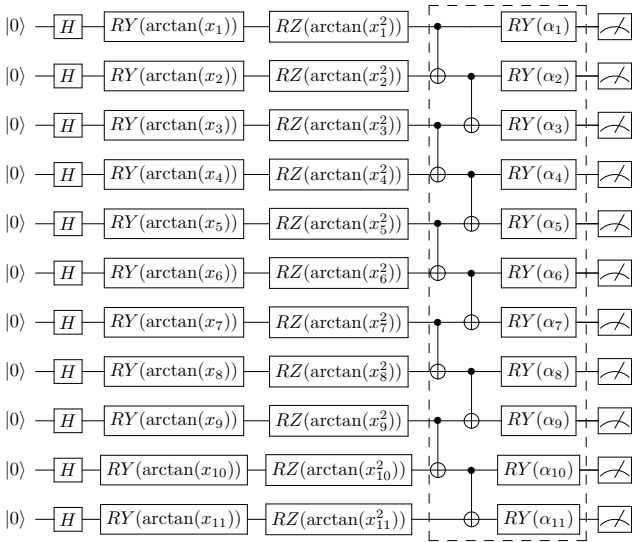

Each LSTM gate (forget, input, output, candidate) uses a separate QNN module that maps the processed classical features to quantum outputs:

\begin{equation}
    \mathbf{f}_t = \text{QNN}_{\theta_f}(\text{NN}_{\omega_f}([\mathbf{x}_t; \mathbf{h}_{t-1}])),
\end{equation}
where $\theta_f$ represents the trainable parameters (rotation angles) of QNN and $\omega_f$ represents the trainable weights in that particular pre-processing NN.
Each gate QNN uses $n$ qubits, initialized in $\ketzero$, with input encoded via angle encoding:
\begin{equation}
    U_{\text{enc}}(\mathbf{x}) = \bigotimes_{i=1}^{n} \RZ{\arctan(x_{i}^2)} \RY{\arctan(x_i)} \text{H},
\end{equation}
where the Hadamard gate $\text{H}$ is used to create unbiased initial state,
\begin{equation}
    \left( H\ket{0}\right)^{\otimes n} = \sum_{(q_1,q_2,...,q_n) \in \{ 0,1\}^n} \frac{1}{\sqrt{2^n}} \ket{q_1} \otimes \ket{q_2} \otimes \cdots \otimes \ket{q_n}.
\end{equation}
The encoded quantum state then goes through the trainable or variational circuit component, which includes multiple $\text{RY}$ gates and \CNOT gates. This trainable circuit block is illustrated in the dashed-line box of Fig.\ref{Fig:QLSTM_VQC_CIRCUIT}. In this paper, we repeat the box 5 times so that there are $11 \times 5 = 55$ trainable parameters in each QNN.
%
Gate values are obtained from expectation values of the $Z$ operator (here we measure first $4$ qubits to construct the hidden state). The cell state $\mathbf{c}_t$ and hidden state $\mathbf{h}_t$ follow standard LSTM rules. Cell state and hidden state are updated as:
\begin{equation}
\mathbf{c}_t = \mathbf{f}_t \odot \mathbf{c}_{t-1} + \mathbf{i}_t \odot \mathbf{g}_t,\quad
\mathbf{h}_t = \mathbf{o}_t \odot \tanh(\mathbf{c}_t).
\end{equation}

\section{Experiments}

\subsection{Dataset and Setup}
We evaluate our models on the ETTh1 dataset \cite{zhou2021informer}, a benchmark multivariate time series dataset from the ETT family, containing hourly data of electricity transformer temperatures with 7 input features. We follow the common forecasting setting: given a past sequence of 16 time steps, the model predicts the next single time step.

All models are trained using the same hyperparameters: 30 epochs, batch size 64, sequence length 16, and Adam optimizer with learning rate $10^{-3}$. For differential privacy training, we adopt Gaussian noise injection with $\ell_2$ clipping norm of 1.0 and vary the noise multiplier $\sigma \in \{0.5, 1.0, 2.0\}$. We repeat each training setup with and without DP enabled.

\subsection{Models and Configurations}
We compare the following models:
\begin{itemize}
    \item \textbf{Baseline (LSTM):} A standard stacked LSTM with 2 layers and hidden size 64.
    \item \textbf{QRWKV:} Our quantum receptance weighted key-value model with 4-qubits.
    \item \textbf{QASA:} The quantum variational self-attention model with 9-qubits.
    \item \textbf{QLSTM:} Quantum-enhanced version of LSTM integrating classical neural networks with quantum neural networks (QNNs). Each QNN operates on 11 qubits. At each time step, the concatenated vector $[\mathbf{x}_t; \mathbf{h}_{t-1}] \in \mathbb{R}^{11}$ is passed through a linear layer with input and output dimensions both equal to 11 (i.e., \texttt{nn.Linear(11, 11)}). The resulting transformed vector is then fed into the QNN for quantum processing.
\end{itemize}

Each quantum model is trained in both private and non-private settings. Experiments were conducted using simulated quantum backends, as specified in Section III.

\section{Results}

\subsection{Forecasting Accuracy Under Privacy Constraints}

Table~\ref{tab:results} reports the MAE, MSE, and RMSE of each model on the ETTh1 dataset under different noise multipliers.

\begin{table}[h]
\caption{Forecasting performance on ETTh1 (seq\_len=16, pred\_len=1)}
\label{tab:results}
\centering
\begin{tabular}{|l|c|c|c|}
\hline
\textbf{Model} & \textbf{MAE} & \textbf{MSE} & \textbf{RMSE} \\
\hline
\multicolumn{4}{|c|}{\textit{Baseline (LSTM)}} \\
\hline
No DP & 0.2539 & 0.1208 & 0.3476 \\
DP ($\sigma$=0.5) & 0.2905 & 0.2238 & 0.4730 \\
DP ($\sigma$=1.0) & 0.3081 & 0.2389 & 0.4887 \\
DP ($\sigma$=2.0) & 0.3370 & 0.2693 & 0.5189 \\
\hline
\multicolumn{4}{|c|}{\textit{QRWKV}} \\
\hline
No DP & 0.2600 & 0.1101 & 0.3317 \\
DP ($\sigma$=0.5) & 0.2694 & 0.1212 & 0.3481 \\
DP ($\sigma$=1.0) & 0.2987 & 0.1753 & 0.4187 \\
DP ($\sigma$=2.0) & 0.3192 & 0.1980 & 0.4448 \\
\hline
\multicolumn{4}{|c|}{\textit{QASA}} \\
\hline
No DP & 0.2415 & 0.1065 & 0.3263 \\
DP ($\sigma$=0.5) & 0.2603 & 0.1324 & 0.3639 \\
DP ($\sigma$=1.0) & 0.2809 & 0.1461 & 0.3824 \\
DP ($\sigma$=2.0) & 0.3037 & 0.1766 & 0.4201 \\
\hline
\multicolumn{4}{|c|}{\textit{QLSTM}} \\
\hline
No DP & 0.3065 & 0.1618 & 0.4022 \\
DP ($\sigma$=0.5) & 0.2971 & 0.4709 & 0.6862 \\
DP ($\sigma$=1.0) & 0.3009 & 0.4045 & 0.6360 \\
DP ($\sigma$=2.0) & 0.3233 & 0.2433 & 0.4933 \\
\hline
\end{tabular}
\end{table}

\subsection{Analysis}

From the results, we observe that all models exhibit a degradation in performance as the noise multiplier increases, which is expected under stronger privacy budgets. Notably, the quantum models (QRWKV and QASA) maintain lower error metrics compared to the classical LSTM baseline across all DP settings. In particular, QASA demonstrates the best MAE and RMSE in both private and non-private regimes, highlighting its robustness to noise perturbation due to the expressiveness of quantum self-attention. This findings on multivariate dataset are the same as \cite{chen2025qrltw}. The QLSTM model maintains similar performance while using a much smaller model size (number of trainable parameters).

These results support our central hypothesis: \textit{variational quantum models can preserve competitive accuracy under strict privacy constraints}, making them suitable for regulated domains such as healthcare and energy analytics.

\section{Conclusion}

We have introduced Q-DPTS, the first framework for differentially private time series forecasting using variational quantum circuits. Our method integrates classical DP-SGD with hybrid quantum models, allowing for privacy-preserving learning without sacrificing predictive accuracy. Through extensive experiments on the ETTh1 dataset, we evaluated the performance of four models under varying privacy budgets, included baseline LSTM, QRWKV, QLSTM, and QASA. The results demonstrate that quantum models, particularly QASA and QRWKV, maintain strong forecasting performance even with high noise multipliers, outperforming classical baselines in both accuracy and robustness.

Our work highlights the feasibility and effectiveness of bringing together differential privacy and quantum machine learning, opening new directions for deploying QML models in sensitive application domains.

\section{Limitations and Future Work}

Despite promising results, this work presents several limitations that warrant further exploration:

\begin{itemize}
    \item \textbf{Quantum Simulation Only:} All experiments used noiseless hardware simulation backends; this does not imply noiseless datasets. Future work will add realistic noise channels (depolarizing, amplitude damping) and real hardware. Future studies should evaluate Q-DPTS on real quantum hardware, where gate noise and qubit decoherence may impact performance and privacy guarantees.
    
    \item \textbf{Scalability Constraints:} The current implementations are limited qubit circuits due to simulation overhead. Scaling to larger models will require architectural optimization and circuit compression.
    
    \item \textbf{Privacy Accounting:} Although we use Rényi Differential Privacy (RDP) for budget tracking, future work should investigate tighter accounting methods or quantum-native privacy frameworks that consider quantum information-theoretic leakage.
    
    \item \textbf{Limited Forecasting Horizon:} We forecast only a single time step ahead. Extending the models to multi-step forecasting and irregular time series would improve practical applicability.
    
    \item \textbf{Application Diversity:} The evaluation is confined to the ETTh1 dataset. Broader benchmarking on financial, biomedical, or human behavioral datasets is essential to validate generalizability.
\end{itemize}

In future work, we also aim to explore quantum federated learning under DP, employ adaptive noise scheduling for utility preservation, and integrate quantum generative models for time series imputation and simulation.

\bibliographystyle{ieeetr}
\bibliography{ref}

\end{document}